\documentclass[runningheads]{llncs}
 \usepackage{times}
 \usepackage{srcltx}
 \usepackage{braket}
 \usepackage{tikz}
 \usepackage{verbatim}
 \usepackage{longtable}
 \usepackage{color}
 \usepackage{stmaryrd} 
 \usepackage{xspace}
 \usepackage{multirow}

\usepackage{amsmath}
\usepackage{latexsym}
\usepackage{textcomp}

\usepackage[bitstream-charter, greekuppercase=italicized, cal=cmcal]{mathdesign}
\usepackage{beramono}
\usepackage[normalem]{ulem}

\usepackage{textcmds}

\usepackage[british]{babel}
\usepackage[latin1]{inputenc}
\usepackage{xcolor}

\usepackage{url}
\urldef{\mailsa}\path|cdpu@dtu.dk|

\newcommand{\is}{:-}
\newcommand{\Request}{\texttt{<Request>}\xspace}
\newcommand{\PS}{\texttt{<PolicySet>}\xspace}
\newcommand{\Policy}{\texttt{<Policy>}\xspace}
\newcommand{\Rule}{\texttt{<Rule>}\xspace}
\newcommand{\Target}{\texttt{<Target>}\xspace}
\newcommand{\Condition}{\texttt{<Condition>}\xspace}
\newcommand{\Match}{\texttt{<Match>}\xspace}
\newcommand{\AnyOf}{\texttt{<AnyOf>}\xspace}
\newcommand{\AllOf}{\texttt{<AllOf>}\xspace}

\newcommand{\eval}{\ensuremath{\mathsf{eval}}\xspace}
\newcommand{\seq}[1]{<\!\!\!< #1 >\!\!\!>}

\newcommand{\prog}[1]{\[{\begin{array}{ll} #1 \end{array} }\] }

\newcommand{\semantics}[1]{\ensuremath{\llbracket #1\rrbracket}\xspace}

\newcommand{\la}{\leftarrow}
\newcommand{\ra}{\rightarrow}

\newcommand{\ground}[1]{\ensuremath{\mathit{ground}(#1)}}
\newcommand{\mc}[1]{\ensuremath{\mathcal{#1}}}
\newcommand{\mb}[1]{\ensuremath{\mathbf{#1}}}

\newcommand{\val}{\ensuremath{\mathsf{val}}}

\newcommand{\m}{\ensuremath{\mathsf{m}}}
\newcommand{\nm}{\ensuremath{\mathsf{nm}}}
\newcommand{\idt}{\ensuremath{\mathsf{idt}}}
\newcommand{\error}{\ensuremath{\mathsf{error}}}
\renewcommand{\t}{\ensuremath{\mathsf{t}}}
\newcommand{\f}{\ensuremath{\mathsf{f}}}
\newcommand{\p}{\ensuremath{\mathsf{p}}}
\renewcommand{\d}{\ensuremath{\mathsf{d}}}
\newcommand{\na}{\ensuremath{\mathsf{na}}}
\newcommand{\dec}{\ensuremath{\mathsf{dec}}}
\newcommand{\comb}{\ensuremath{\mathsf{CombID}}}

\newcommand{\algo}{\ensuremath{\mathsf{algo}}}
\newcommand{\po}{\ensuremath{\mathsf{po}}}
\newcommand{\denyo}{\ensuremath{\mathsf{do}}}
\newcommand{\ooa}{\ensuremath{\mathsf{ooa}}}
\newcommand{\fa}{\ensuremath{\mathsf{fa}}}
\newcommand{\mynot}{\ensuremath{\mathbf{not}~}}

\newcommand{\cond}{\ensuremath{\mathit{cond}}}

\newcommand{\gap}{\ensuremath{\mathit{gap}}}
\newcommand{\conflict}{\ensuremath{\mathit{conflict}}}
\newcommand{\reachable}{\ensuremath{\mathit{reachable}}}
\newcommand{\notreachable}{\ensuremath{\mathit{not\_reachable}}}

\begin{document}


\title{Detecting Incompleteness, Conflicting and Unreachability XACML Policies using Answer Set Programming}
%
\titlerunning{Detecting Incompleteness, Conflicting  and Unreachability XACML Policies using ASP}

\author{Carroline Dewi Puspa Kencana Ramli}
\authorrunning{Carroline Dewi Puspa Kencana Ramli}

\institute{Department of Informatics and Mathematical Modelling \\
Danmarks Tekniske Universitet \\
Lyngby, Denmark\\
\email{\mailsa}}

\maketitle
\vspace{-15pt}
\begin{abstract}
Recently, XACML is a popular access control policy language that is used widely in many applications. Policies in XACML are built based on many components over distributed resources. Due to  the expressiveness of XACML, it is not trivial for policy administrators to understand the overall effect and consequences of XACML policies they have written. In this paper we show a mechanism and a tool how to analyses big access control policies sets such as (i) incompleteness policies, (ii) conflicting policies, and (iii) unreachable policies. To detect these problems we present a method using Answer Set Programming (ASP) in the context of XACML 3.0.  

\end{abstract}


\section{Introduction}
XACML (eXtensible Access Control Markup Language) is an OASIS\footnote{OASIS (Organization for the Advancement of Structured Information Standard) is a non-for-profit,
global consortium that drives the development, convergence, and adoption of e-business standards. Information about OASIS can be found at \url{http://http://www.oasis-open.org/}.} standard that describes both a policy language and a query/response language for access control policies. It has been used in many different applications range over health care information systems, transport systems to banking information systems\footnote{XACML references and products can be seen in \url{https://www.oasis-open.org/committees/download.php/42588/xacmlRefs-V1-85.html}.}.  The policy language is used to express access control requirements (\textit{who can access what when}) over distributed resources and  the query/response language is used to query whether a particular access should be allowed (\textit{request}) and to answer the query (\textit{response}). Access control policies in XACML are built based on many components and combined using a particular combining algorithm.   

Due to  the expressiveness of XACML, it is not trivial for policy administrators to understand the overall effect and consequences of XACML policies they have written. The problem becomes more prevalent if there are no mechanisms/automated tools to analyse big chunk of policies. Several problems might occur in developing access control policies such as \textit{incomplete policies} and \textit{conflicting policies}. Moreover, detecting \textit{unreachable policies} might help policy administrators to remove unused policies in order to make the set of policies slimmer and make it easier to be maintained. 

\paragraph{Analysing Incomplete Policies.} It is high probable that policy developers do not define all possible situations that might occur. Incomplete access control policies might lead  to a security problem. Following we present a probable scenario how an intruder can use this security hole to get an access.

In XACML, PDP (Policy Decision Point) computes a decision based on administrated policies in a database, but the final decision is made by PEP (Policy Enforcement Point). There are two PEP-biased:
\begin{enumerate}
\item Permit-biased PEP: if the decision from PDP is deny, then the PEP shall deny assess. All other decisions shall result in the permission of access. 
\item Deny-biased PEP: if the decision from PDP is permit, then the PEP shall permit the access. All other decisions shall result in the denial of access. 
\end{enumerate}
In this case, there is a possibility that an intruder can get an access unintentionally by trying to query so that the response is no policy is applicable. Using Permit-biased PEP, the decision will let the intruder have access to the system. 

\paragraph{Analysing Conflicting Policies.} Conflicting policies can have serious consequences and may lead to unauthorized access. Basically, in XACML, conflicting decision never occurs since all policies are combined with a particular combining algorithm that only returns one decision. However it is interesting to analyse conflict in between  policies for example different department can have different decision. By analysing conflict, the policy makers can rethink again whether they made correct policies. 

\paragraph{Analysing Unreachable Policies.} Analysisng unreachable policies  helps policy administrators to reduce the size of the set of policies. A policy is unreachable if for all request it never gives decision i.e., either it always not applicable or there is another policy that overrides its decision. It is safe to remove unreachable policies because their decisions never influence the final decisions. 

\paragraph{}To address the above concern we propose a logic-based XACML analysis framework using logic programs (LPs) and  answer set  semantics.  Answer Set Programming (ASP) has become a popular approach to solve combinatorial problems declaratively. There are several efficient implementations of answer set solvers, such as \texttt{ASSAT}\footnote{\url{http://assat.cs.ust.hk/}, }, \texttt{clasp}\footnote{\url{http://www.cs.uni-potsdam.de/clasp/}}, \texttt{Cmodels}\footnote{\url{http://www.cs.utexas.edu/users/tag/cmodels/}}, \texttt{Smodels}\footnote{\url{http://www.tcs.hut.fi/Software/smodels/}}, and many more. 
We present in this paper a method using ASP to solve those problems explained previously in the context of XACML 3.0 \cite{XACML3.0}, the most recent version of XACML. 


\paragraph{Outline.}  In this paper first we explain the model and semantics of XACML 3.0 in Sect.~\ref{s:xacml}. Then we describe the mapping of XACML 3.0 components into logic programs $\mc{P}_{XACML}$ in Sect.~\ref{s:mapping}. 
Next we show how to analyse access control policies such as incompleteness, conflicting and reachability XACML policies in Sect.~\ref{s:analysis}. We end the paper with conclusion and future work. 

\section{XACML Model and Semantics}
\label{s:xacml}
In this section we briefly describe the XACML policy language and XACML query language model. First we show the faithfully abstract syntax XACML 3.0.  Then we present a   semantics of XACML 3.0 without considering indeterminate values. Our argument is that we evaluate access control properties to a set of policies in statically. Hence,  indeterminate values which only occur when there are errors during evaluation process do not give impact to our analysis. 
At the end of this section we show the semantics of XACML combining algorithms which are used for composing several access control policies.

\subsection{Abstract Syntax of XACML 3.0}
We summarize the syntax of XACML 3.0 in Table  \ref{t:syntax}. To make the notation clear we use bold font for non-terminal \textbf{symbols}, typewriter font for terminal \texttt{symbols} and \textit{identifiers} and \textit{values} are written in italic font. Moreover, \texttt{<XACML Component>} denotes the symbol for XACML component. We use the star symbol (*) to indicate that there is zero or more of the preceding element and we use the plus symbol ($^+$) to indicate that there is one or more of the preceding element. We assume that each policy must have a unique identifier (ID).

\begin{table*}[ht!]
\vspace{-10pt}
\caption{Abstraction of XACML 3.0 Components}
\label{t:syntax}
\vspace{-15pt}
\begin{center}
\begin{tabular}{|lcl|}
\hline  \vspace{-7pt} & & \\ 
\multicolumn{3}{|c|}{\textbf{\underline{XACML Policy Components}}} \\ \vspace{-7pt}& &\\
\PS & \is &  \textit{PolicySetID} \texttt{= [}\Target ,\seq{PolicySetID*}, \textbf{CombID} \texttt{]}\\
            & \textbar &  \textit{PolicySetID} \texttt{= [}\Target, \seq{\textit{PolicyID*}}, \textbf{CombID} \texttt{]}\\
\Policy & \is &  \textit{PolicyID} \texttt{= [}\Target, \seq{\textit{PolicySetID$^+$}} \textbf{CombID} \texttt{]}\\
\Rule      & \is &  \textit{RuleID} \texttt{= [} \textbf{Effect}\texttt{,} \Target \texttt{,} \Condition \texttt{]}\\
\Condition  & \is & \textit{propositional formulae} \\
\Target & \is & $\texttt{Null}$  \\
             & \textbar & $\bigwedge$ \AnyOf$^+$ \\
\AnyOf & \is & $\bigvee$ \AllOf$^+$ \\
\AllOf & \is & $\bigwedge$ \Match$^+$ \\
\Match & \is & \textbf{AttrType}\texttt{(} \textit{attribute value} \texttt{)} \\
\textbf{CombID}  & \is & \texttt{po} \textbar\ \texttt{do} \textbar\ \texttt{fa} \textbar\ \texttt{ooa} \\
\textbf{Effect} & \is & \texttt{deny} \textbar\ \texttt{permit}  \\ 
\textbf{AttrType}& \is & \texttt{subject} \textbar\ \texttt{action} \textbar\ \texttt{resource} \textbar\ \texttt{environment} \\
\hline \vspace{-7pt} & & \\ 
\multicolumn{3}{|c|}{\textbf{\underline{XACML Request Component}}}\\ \vspace{-7pt}& &\\
\Request & \is & \texttt{\{} \textbf{Attribute$^+$}\texttt{\}}\\
\textbf{Attribute}			& \is &\textbf{AttrType}\texttt{(} \textit{attribute value} \texttt{)} \textbar\  
\textit{external state} \\
\hline
   \end{tabular}
\vspace{-30pt}
\end{center}
\end{table*}

There are three levels of policies in XACML, namely \PS, \Policy and \Rule. \PS or \Policy can act as the root of a set of access control policies while \Rule is a single entity that describes one particular access control policy. Through this paper, we assume that \PS is the root of the set of access control policies.

\PS and \Policy have the same characteristic, i.e., they are containers for a sequence of  \PS, \Policy or \Rule. A \PS  contains either a sequence of  \PS or a sequence of \Policy while a \Policy only can contains a sequence of \Rule. The sequence of \PS, \Policy or \Rule is combined with a particular \emph{combining algorithm}. There are four common combining algorithms defined in XACML 3.0, namely \emph{permit-overrides} ($\po$), \emph{deny-overrides} ($\denyo$), \emph{first-applicable} ($\fa$) and \emph{only-one-applicable} ($\ooa$). 

A  \Rule describes an individual access control policy. It regulates whether an access should be \textit{permitted} or \textit{denied}. All \PS, \Policy and \Rule are applicable whenever their \Target matches with the \Request. When the \Rule's \Target matches with the \Request, then the applicability of the \Rule is refined by its \Condition. 

A \Target is a combination of \Match elements. Each \Match element describes an \emph{attribute} that a \Request should match in order to activate a policy. There are four attribute categories in XACML 3.0, namely \emph{subject} attribute, \emph{action} attribute, \emph{resource} attribute and \emph{environment} attribute. The subject attribute is the entity requesting access, e.g., a file system, a workstation, etc. The action attribute defines the type of access requested, e.g., to read, to write, to delete, etc. The resource attribute is a data, service or system components. The environment attribute can optionally provide additional information. 

A \Request contains a set of attributes information about  access request. A \Request can contain additional information such as external state condition (e.g. the current time, current temperature, etc).

\subsection{XACML 3.0 Formal Semantics}
\label{ss:formal semantics}
The evaluation of XACML policies against a given request starts from the evaluation of \Match elements and continued bottom-up until the evaluation of \PS as the root element. We use the \semantics{.} notation to map XACML elements into their values (see the summary in Tabel \ref{t:xacml values}). 

\begin{table}
\vspace{-15pt}
\caption{XACML Components' Values}
\label{t:xacml values}
\begin{center}
\begin{tabular}{|p{0.3\textwidth}|l|}
\hline
\multicolumn{1}{|c}{\textbf{XACML Components} }&\multicolumn{1}{|c|}{ \textbf{Values}}\\
\hline
\semantics{\Match}, \semantics{\AllOf}, \semantics{\AnyOf}, \semantics{\Target}& match (\m) and not match (\nm) \\
\hline
\semantics{\Condition} & true (\t) and false (\f) \\
\hline
\semantics{\Rule}, \semantics{\Policy}, \semantics{\PS}& permit (\p), deny (\d) and not applicable (\na) \\
\hline
\end{tabular}
\vspace{-35pt}
\end{center}
\end{table}

\vspace{5pt}
\noindent\textbf{Evaluation of \Match into $\Set{\m, \nm}$.}
Given a \Request $Q$, the evaluation of \Match $M$ is as follows
\begin{equation}
\label{eq:match}
   \semantics{M}(Q) = 
   \begin{cases}
  \m & \textrm{if } M \in Q \\
  \nm & \textrm{if }M \not\in Q \\
    \end{cases}
\end{equation}

\noindent\textbf{Evaluation of \AllOf into $\Set{\m, \nm}$.}
Given a \Request $Q$, the evaluation of \AllOf $A = \bigwedge_{i = 1}^n M_i$ is as follows
\begin{equation}
\label{eq:allof}
   \semantics{A}(Q) = 
   \begin{cases}
  \m   & \textrm{if } \forall i:   \semantics{M_i} = \m\\
  \nm & \textrm{if } \exists i: \semantics{M_i} = \nm \\
    \end{cases}
\end{equation}
where each $M_i$ is a \Match element. 

\vspace{3pt}
\noindent\textbf{Evaluation of \AnyOf into $\Set{\m, \nm}$.}
Given a \Request $Q$, the evaluation of \AnyOf $E = \bigvee_{i = 1}^n A_i$ is as follows
\begin{equation}
\label{eq:anyof}
   \semantics{E}(Q) = 
   \begin{cases}
  \m   & \textrm{if } \exists i: \semantics{A_i} = \m\\
  \nm & \textrm{if } \forall   i: \semantics{A_i} = \nm \\
    \end{cases}
\end{equation}
where each $A_i$ is a \AllOf element. 

\vspace{5pt}
\noindent\textbf{Evaluation of \Target into $\Set{\m, \nm}$.}
Given a \Request $Q$, the evaluation of \Target $T = \bigwedge_{i = 1}^n E_i$ is as follows
\begin{equation}
\label{eq:target}
   \semantics{T}(Q) = 
   \begin{cases}
  \m   & \textrm{if } \forall i:   \semantics{E_i} = \m \mbox{ or } T = \texttt{Null}\\
  \nm & \textrm{if } \exists i: \semantics{E_i} = \nm \\
    \end{cases}
\end{equation}
where each $E_i$ is a \AnyOf element. 
Empty \Target, indicated by \texttt{Null} always evaluated to \m. 

\vspace{5pt}
\noindent\textbf{Evaluation of \Condition into $\Set{\t, \f}$.}
Given a \Request $Q$, the evaluation of \Condition $C$ is as follows
\begin{equation}
\label{eq:condition}
   \semantics{C}(Q) =  \eval(C, Q)
\end{equation}
\noindent \textit{Note:} The \eval is an unspecified function that returns $\Set{\t, \f}$. 

\vspace{5pt}
\noindent\textbf{Evaluation of \Rule into $\Set{\d, \p, \na}$.}
Given a \Request $Q$, the evaluation of \Rule $R = [e, T, C]$ as follows
\begin{equation}
\label{eq:rule}
   \semantics{R}(Q) = 
   \begin{cases}
  e   & \textrm{if } \semantics{T}(Q) = \m \mbox{ and } \semantics{C}(Q) = \t\\
  \na& \textrm{if } (\semantics{T}(Q) = \m \mbox{ and } \semantics{C}(Q) = \f)  \mbox{ or} \semantics{T}(Q) = \nm \\
    \end{cases}
\end{equation}
where $e \in \Set{\p, \d}$,  $T$ is a \Target element and $C$ is a \Condition element.

\vspace{5pt}
\noindent\textbf{Evaluation of \Policy into $\Set{\d, \p, \na}$.}
Given a \Request $Q$, the evaluation of \Policy $P = [T, \seq{R_1, \ldots, R_n}, \comb]$ is as follows

\begin{equation}
\label{eq:policy}
   \semantics{P}(Q) = 
   \begin{cases}
  \na& \textrm{if } \semantics{T}(Q) = \nm \mbox{ or } \forall i: \semantics{R_i}(Q) = \na\\
  \bigoplus_{\comb}(\mb{R})& \mbox{otherwise}
    \end{cases}
\end{equation}
where $T$ is a \Target element, and each $R_i$ is a \Rule element. We use $\mb{R}$ to denote $\seq{\semantics{R_1}(Q), \ldots, \semantics{R_n}(Q)}$.

 \noindent \textit{Note:} The combining algorithm denoted by $\bigoplus_{\comb}$ will be explained in Sect. \ref{ss:combining algorithms}.

\vspace{5pt}
\noindent\textbf{Evaluation of \PS into $\Set{\d, \p, \na}$.}
Given a \Request $Q$, the evaluation of \PS $PS = [T, \seq{P_1, \ldots, P_n}, \comb]$ is as follows

\begin{equation}
\label{eq:ps}
   \semantics{PS}(Q) = 
   \begin{cases}
  \na& \textrm{if } \semantics{T}(Q) = \nm \mbox{ or } \forall i: \semantics{P_i}(Q) = \na\\
  \bigoplus_{\comb}(\mb{P})& \mbox{otherwise}
    \end{cases}
\end{equation}
where $T$ is a \Target element and each $P_i$ is a \Policy (or \PS) element. We use $\mb{P}$ to denote $\seq{\semantics{P_1}(Q), \ldots, \semantics{P_n}(Q)} $.

\subsection{XACML Combining Algorithms}
\label{ss:combining algorithms}
There are four common combining algorithms defined in XACML 3.0, namely permit-overrides (\po), deny-overrides (\denyo), first-applicable (\fa) and only-one-applicable (\ooa). The permit-overrides combining algorithm takes permit decision as the most priority than deny decision while the deny-overrides combining algorithm takes deny decision over permit. Likewise their names, the first-applicable combining algorithm return the first \Rule (or \Policy or \PS) that is applicable (either permit (\p) or deny(\d) value) and the only-one-applicable combining algorithm return a decision whenever only one \Rule (or \Policy or \PS) which is applicable, otherwise it returns not applicable (\na).  

\subsubsection{Permit-Overrides}
The permit-overrides combining algorithm is intended for those cases where a permit
decision should have priority over a deny decision.

Let $S = \seq{v_1, v_2, \ldots, v_n}$ be a sequence of policy values. The permit-overrides combining algorithm, $\bigoplus_{\mathtt{po}}$, is defined as follows
\begin{equation}
\label{eq:po}
\bigoplus_{\mathtt{po}} (S) = 
\begin{cases}
\p & \textrm{if $\exists i: v_i = \p$} \\
\d & \textrm{if $\forall i: v_i \neq \p$ and $\exists j: v_j = \d$} \\
\na & \textrm{otherwise}
\end{cases}
\end{equation}

\subsubsection{Deny-Overrides}
The deny-overrides is the mirror of permit-overrides whereas the deny decision has more priority over a permit decision. 

Let $S = \seq{v_1, v_2, \ldots, v_n}$ be a sequence of policy values. The deny-overrides combining algorithm, $\bigoplus_{\mathtt{do}}$, is defined as follows
\begin{equation}
\label{eq:do}
\bigoplus_{\mathtt{do}} (S) = 
\begin{cases}
\d & \textrm{if $\exists i: v_i = \d$} \\
\p & \textrm{if $\forall i: v_i \neq \d$ and $\exists j: v_j = \p$} \\
\na & \textrm{otherwise}
\end{cases}
\end{equation}

\subsubsection{First-Applicable}
The result of first-applicable algorithm is the first Rule, Policy or PolicySet element in the sequence whose is applicable.

Let $S = \seq{v_1, v_2, \ldots, v_n}$ be a sequence of policy values. The first-applicable combining algorithm, $\bigoplus_{\mathtt{fa}}$, is defined as follows
\begin{equation}
\label{eq:fa}
\bigoplus_{\mathtt{fa}} (S) = 
\begin{cases}
v_i & \textrm{if $\exists i: v_i \neq \na$ and $\forall j: j < i \Rightarrow v_j = \na$}\\
\na & \textrm{if $\forall i: v_i = \na $}
\end{cases}
\end{equation}

\subsubsection{Only-One-Applicable}
The result of the only-one-applicable combining algorithm ensures that one and only
one policy is applicable. If no policy applies, then the result is \na, but if more than one policy is applicable, then the result is \idt. When exactly one policy is applicable, the result of the combining algorithm is the result of evaluating the single applicable policy.  Please note that we do not use \idt\ in this step. Hence, all of \idt\ value is converted to \na.  

Let $S = \seq{v_1, v_2, \ldots, v_n}$ be a sequence of policy values. The only-one-applicable  combining algorithm, $\bigoplus_{\mathtt{ooa}}$, is defined as follows
\begin{equation}
\label{eq:ooa}
\bigoplus_{\mathtt{ooa}} (S) = 
\begin{cases}
v_i & \textrm{if $\exists i: v_i \neq \na$ and $\forall j: j \neq i \Rightarrow v_j = \na$}\\
\na & \textrm{if ($\exists i, j: i \neq j$ and $v_i \neq \na$ and $v_j \neq \na$) or} \\
& \textrm{if $\forall i: v_i = \na $}
\end{cases}
\end{equation}

\section{Mapping XACML into Logic Programs}
\label{s:mapping}
First we explain the syntax of logic program (LP) in this section. Then we show the transforming XACML 3.0 into LP. The semantics of LP is explained in the following section when we use it for analysis purposes. 

\subsection{Syntax of Logic Programs}
We start by introducing some notations and terminologies which we will use through the paper. 
 \vspace{5pt}
 \noindent\textbf{First-Order Language.}
 We consider an \emph{alphabet} consisting of (finite or countably infinite) disjoint sets of variables, constants, function symbols, predicate symbols, connectives $\Set{\mathbf{not}, \wedge, \la}$,  punctuation symbols $\Set{"(", ",", ")", "."}$ and special symbols $\Set{\top, \bot}$. In this paper we will use upper case letters to denote variables and lower case letters to denote constants, function and predicate symbols. Terms, atoms, literals and formulae are defined as usual. The \emph{language} given by an alphabet  consists of the set of all formulae constructed from the symbols occurring in the alphabet.
 
 \noindent\textbf{Logic Programs.}
 A \emph{rule}  is an expression of the form 
\begin{equation}\label{rule} A \la B_1 \wedge \ldots \wedge B_m \wedge \mynot B_{m+1} \wedge \ldots \wedge \mynot B_n. \end{equation} 
where $A$ is either an atom or $\bot$ and each $B_i$, $1 \leq i \leq n$, is an atom or $\top$. $\top$ is a valid formula. 
$A$ is called the \emph{head} and $B_1 \wedge \ldots \wedge B_m \wedge \mynot B_{m+1} \wedge \ldots \wedge \mynot B_n$ is called \emph{body} of the clause. 
We usually write $B_1 \wedge \ldots \wedge B_m \wedge \mynot B_{m+1} \wedge \ldots \wedge \mynot B_n$ simply as $B_1, \ldots, B_m, \mynot B_{m+1}, \ldots, \mynot B_n$.  

We refer the rule as a \emph{constraint} when $A$ is  $\bot$. One should observe that the body of a rule must not be empty. A rule of the form $A \la \top$ is called a \emph{fact}. 
 
 A \emph{logic program} (LP) is a finite set of rules. \ground{\mc{P}} denotes the set of all ground instances of the program \mc{P}.
 
\subsection{XACML Transformations}
The transformation of XACML components is based on the semantics of each component explained in Sect. \ref{ss:formal semantics}.  First we recall the syntax of each component then we show how the transformation is. 

\vspace{5pt}
\noindent\textbf{\Request Transformation.} \textit{XACML Syntax:} Let $Q = \Set{A_1, \ldots, A_n}, 1 \leq i \leq n$, be a \Request component. The transformation of \Request, $Q$, into LP $\mc{P}_{Q}$ is as follows
\[ A_i \la \top. \ \ 1 \leq i \leq n \]

\noindent\textbf{\Match Transformation.}
\textit{XACML Syntax:} Let $M$ be a \Match component. The transformation of \Match $M$  into LP $\mc{P}_{M}$ is as follows (see \eqref{eq:match} for \Match evaluation)
\prog{
\val(M, \m)  & \la M. \\
\val(M, \nm)& \la \mynot M. \\
}

\noindent\textbf{\AllOf Transformation.}
\textit{XACML Syntax:} Let $A =\bigwedge_{i=1}^n M_i$ be an \AllOf component where each $M_i$ is a \Match component. The transformation of \AllOf $A$  into LP $\mc{P}_{A}$ is as follows (see \eqref{eq:allof} for \AllOf evaluation)
\prog{
\val(A, \m)   & \la \val(M_1, \m), \ldots, \val(M_n, \m).  \\
\val(A, \nm) & \la \val(M_i, \nm).\  (1 \leq i \leq n) \\
}

\noindent\textbf{\AnyOf Transformation.}
\textit{XACML Syntax:} Let $E = \bigvee_{i = 1}^{n} A_i$ be an \AnyOf component where each $A_i$ is an \AllOf component. The transformation of \AnyOf $E$  into LP $\mc{P}_{E}$ is as follows (see \eqref{eq:anyof} for \AnyOf evaluation)
\prog{
\val(E, \m) & \la \val(A_i, \m). \ (1 \leq i \leq n) \\
\val(E, \nm)   & \la \val(A_1, \nm), \ldots, \val(A_n, \nm).  \\
}

\noindent\textbf{\Target Transformation.}
\textit{XACML Syntax:} Let $T = \bigwedge_{i=1}^{n} E_i$ be a \Target component where each $E_i$ is an \AnyOf component. The transformation of \Target $T$  into LP $\mc{P}_{T}$ is as follows (see \eqref{eq:target} for \Target evaluation)
\prog{
\val(T, \m)   & \la \val(E_1, \m), \ldots, \val(E_n, \m).  \\
\val(null, \m) & \la \top. \\
\val(T, \nm) & \la \val(E_i, \nm).\  (1 \leq i \leq n) \\
}

\noindent\textbf{\Condition Transformation.}
\textit{XACML Syntax:}  We assume that the \Condition element is a boolean formula which the evaluation of \Condition is based on \eval function. The transformation of \Condition $C$ into LP $\mc{P}_{C}$ is as follows
\prog{
\val(C, V) & \la \eval(C, V). 
}
In our previous example of \Rule $r1$, the \Condition$\cond(r1)$ is 
\verb+patient.id(X)+ \verb+/\ patient_record.id(X)+. The possibility of \eval function is like following
\prog{\mc{P}_{\cond(r1)}:\\
\val(\cond(r1), V) & \la  \eval(\cond(r1), V). \\
\eval(\cond(r1), \t) & \la patient\_id(X), patient\_record\_id(X).\\
\eval(\cond(r1), \f) & \la patient\_id(X), patient\_record\_id(Y), X \neq Y.
}
The $\error(patient\_id(X))$ and $\error(patient\_record\_id(X))$ indicate possible errors might occur, e.g., the system could not connect to the database so that the system does not know the ID of the patient. 

\vspace{5pt}
\noindent\textbf{\Rule Transformation.}
\textit{XACML Syntax:} Let $R_{id} = [E, T, C]$ be a \Rule component where $E \in \Set{\p, \d}$, $T$ is a \Target and $C$ is a \Condition.  The transformation of \Rule $R_{id}$ into LP $\mc{P}_{R_{id}}$ is as follows (see \eqref{eq:rule} for \Rule evaluation)
\prog{
\val(R_{id}, E) & \la \val(T, \m), \val(C, \t). \\
\val(R_{id}, \na) & \la \val(T, \m), \val(C, \f). \\
\val(R_{id}, \na) & \la \val(T, \nm). \\
}

\noindent\textbf{\Policy Transformation.}
\textit{XACML Syntax:} Let $P_{id} = [T, \seq{R_1, \ldots, R_n}, \comb]$ be a \Policy component where $T$ is a \Target, $<R_1, \ldots, R_n>$ be a sequence of \Rule elements and $\comb$ be a combining algorithm identifier. 
In order to indicate that the \Policy contains \Rule $R_i$, thus for every \Rule $R_i$ contained in $P_{id} = [T, \seq{R_1, \ldots, R_n}, \comb]$, $\mc{P}_{P_{id}}$ also contains:
\prog{
\dec(P_{id}, R_i, E) & \la \val(R_i, E). \ (1 \leq i \leq n)
}
Next, we do a transformation for \Policy $P_{id}$ and add into LP $\mc{P}_{P_{id}}$ is as follows (see \eqref{eq:policy} for \Policy evaluation)
\prog{
\val(P_{id}, \na) & \la \val(T, \nm). \\
\val(P_{id}, \na) & \la \val(R_1, \na), \ldots, \val(R_n, \na). \\
\val(P_{id}, E) & \la \val(T, \m), \dec(P_{id}, R, V), V \neq \na, \algo(\comb, P_{id}, E). \\
}
We write formulae $\dec(P_{id}, R, V), V \neq \na$ to make sure that there is a \Rule in the \Policy that is not evaluated to \na. We do this to avoid a return value from a combining algorithm that is not \na even tough all of the \Rule elements are evaluated to \na. 

\vspace{5pt}
\noindent\textbf{\PS Transformation.}
The transformation of \PS is similar to the transformation of \Policy component. 

\textit{XACML Syntax:} Let $PS_{id} = [T, \seq{P_1, \ldots, P_n}, \comb]$ be a \Policy component where $T$ is a \Target, $<P_1, \ldots, P_n>$ be a sequence of \Policy (or \PS) elements and $\comb$ be a combining algorithm identifier. The transformation of \PS $PS_{id}$ into logic program $\mc{PS}_{P_{id}}$ is as follows
For every \Policy (or \PS) contained in $PS_{id} = [T, \seq{P_1, \ldots, P_n}, \comb]$, $\mc{P}_{PS_{id}}$ also contains:
\prog{
\dec(PS_{id}, P_i, E) & \la \val(P_i, E). \ (1 \leq i \leq n)
}
And we following rules into $\mc{P}_{PS_{id}}$ 
\prog{
\val(PS_{id}, \na) & \la \val(T, \nm). \\
\val(PS_{id}, \na) & \la \val(P_1, \na), \ldots, \val(P_n, \na). \\
\val(PS_{id}, E) & \la \val(T, \m), \dec(PS_{id}, P, V), V \neq \na, \algo(\comb, PS_{id}, E).\\
}

\subsection{Combining Algorithm Transformation}
We use  $P$ for an variable of \Policy identifier and $R$, $R_1$ and $R_2$ for  variables of \Rule identifiers. In case the evaluation of \PS, the input $P$ is for \PS identifier, $R, R_1$ and $R_2$ are for \Policy (or \PS) identifiers. 

\vspace{5pt}
\noindent\textbf{Permit-Overrides Transformation.}
Let $\mc{P}_{\po}$ be a LP obtained by permit-overrides combining algorithm transformation (see \eqref{eq:po} for the permit-overrides combining algorithm semantics). $\mc{P}_{\po}$ contains: 
\vspace{-5pt}\prog{
\algo(\po, P, \p) & \la \dec(P, R, \p). \\
\algo(\po, P, \d) & \la \mynot \algo(\po, P, \p), \dec(P, R, \d). \\
\algo(\po, P, \na) & \la  \mynot \algo(\po, P, \p), \mynot \algo(\po, P,\d).
}

\vspace{5pt}
\noindent\textbf{Deny-Overrides Transformation.}
Let $\mc{P}_{\do}$ be a LP obtained by deny-overrides combining algorithm transformation (see \eqref{eq:do} for the permit-overrides combining algorithm semantics). $\mc{P}_{\po}$ contains: 
\vspace{-5pt}\prog{
\algo(\po, P, \d) & \la \dec(P, R, \d). \\
\algo(\po, P, \p) & \la \mynot \algo(\po, P, \d), \dec(P, R, \p). \\
\algo(\po, P, \na) & \la  \mynot \algo(\po, P, \d), \mynot \algo(\po, P,\d).
}

\noindent\textbf{First-Applicable Transformation.}
Let $\mc{P}_{\fa}$ be a logic program obtained by first-applicable combining algorithm transformation (see \eqref{eq:fa} for the first-applicable  combining algorithm semantics). For each \Policy (or \PS) which uses first-applicable combining algorithm, $P_{id} = [T, \seq{R_1, \ldots, R_n}, \fa]$,  $\mc{P}_{P_{id}}$ contains:
\prog{
\algo(\fa, P, E) & \la \dec(P, R_1, E), E \neq \na. \\
\algo(\fa, P, E) & \la \dec(P, R_1, \na), \dec(P, R_2, E), E \neq \na.\\
& \vdots\\
\algo(\fa, P, E) & \la \dec(P, R_1, \na), \ldots, \dec(P, R_{n-1}, \na), \\
& \hspace{4mm} \dec(P, R_n, E).
}
\noindent\textbf{Only-One-Applicable Transformation.}
Let $\mc{P}_{\ooa}$ be a logic program obtained by only-one-applicable combining algorithm transformation (see \eqref{eq:ooa} for the only-one-applicable combining algorithm semantics). $\mc{P}_{\ooa}$ contains:
\prog{
not\_one\_applicable(P) & \la \dec(P, R1, X), \dec(P, R2, Y), R1 \neq R2, X \neq \na, Y \neq \na. \\
\algo(\ooa, P, E) & \la  \dec(P, R, E), \mynot not\_one\_applicable(P). \\ 
\algo(\ooa, P, \na) & \la not\_one\_applicable(P). 
}
\section{Policy Analysis}
\label{s:analysis}
We use the semantics of LP $\mc{P}_{XACML}$ -- the result  from transforming XACML components into series of LPs -- to analyse access control policy properties. In this section, we present three policy analysis cases namely analysing on incompleteness policies, conflicting policies and unreachable policies. The completeness and free of conflict properties have been introduces by Samarati and di Vimercati in \cite{Samarati2001} and formalized using Belnap four-valued logic \cite{Belnap1977} by  Bruns and Huth in \cite{Bruns2008}. In this section we show how we present ASP programs to capture those properties\footnote{We call ASP programs for logic programs with answer set semantics.}. Our intention is to have an automatic tool that  shows XACML formalization and  in the same time it can be used to help policy administrators to analyse their policies sets. 

 \subsection{Semantics of Logic Programs}
 The declarative semantics of a logic program is given by a model-theoretic semantics  of formulae in the underlying language. The formal definition of answer set semantics can be found in many literatures like in \cite{Baral2003,Gelfond2007}. 
 
 \subsubsection{Interpretations and Models}
The \emph{Herbrand Universe} \mc{U_\mc{L}} for a language \mc{L} is the set of all ground terms that can be formed from the constants and function symbols appearing in \mc{L}. The \emph{Herbrand base} \mc{B_{\mc{L}}} for a language \mc{L} is the set of all ground atoms that can be formed by using predicate symbols from \mc{L} and ground terms from \mc{U_{\mc{L}}} as arguments. By \mc{B_{\mc{P}}} we denote the Herbrand base for language underlying the program \mc{P}. When the context is clear, we are safe to omit  \mc{P}. 

An \emph{interpretation} $I$ of a program \mc{P} is a mapping from the Herbrand base \mc{B_{\mc{P}}} to the set of truth value true and false ($\Set{\top, \bot}$). All atoms belong to interpretation $I$ are mapped to $\top$. All atoms which does not occur in $I$ are mapped to $\bot$. 

The truth value of arbitrary formulae under some interpretation can be determined from a truth table as usual (see Table \ref{t:truth table}). 
\begin{table}
\centering
\vspace{-15pt}
\caption{Truth Values for Formulae}
\vspace{-15pt}
\label{t:truth table}
\[
\begin{array}{c|c|c|c|c}
\phi & \psi & \mynot \phi & \phi \wedge \psi& \phi \la \psi  \\
\hline
\top & \top & \bot & \top & \top \\
\top & \bot & \bot & \bot & \top \\
\bot & \top & \top & \bot & \bot \\
\bot & \bot & \top & \bot & \top
\end{array}
\]
\vspace{-15pt}
\end{table}

The logical value of ground formulae can be derived from Table \ref{t:truth table} in the usual way. A formula $\phi$ is then \emph{true under interpretation $I$}, denoted by $I(\phi) = \top$, if all its ground instances are true in $I$; it is \emph{false under interpretation $I$}, denoted by $I(\phi) = \bot$, if there is a ground instance of $\phi$ that is false in $I$. 

Let $I$ be an interpretation. $I$  \emph{satisfies} formula $\phi$, denoted by $I \models \phi$, if $I(\phi) = \top$. For a program \mc{P}, we say $I$ \emph{satisfies} of \mc{P}, denoted by $I \models \mc{P}$,  if $I$ satisfies for every clause in \mc{P}. 

Let \mc{I} be a collection of interpretations. Then an interpretation $I$ is \mc{I} is called \emph{minimal} in \mc{I} if and only if there is no interpretation $J$ in \mc{I} such that $J \subsetneq I$. An interpretation $I$ is called \emph{least} in \mc{I} if and only if $I \subseteq J$ for any interpretation $J$ in \mc{I}. A model $M$ of a program \mc{P} is called minimal (respectively least) if it is minimal (respectively least) among all models of \mc{P}.

The answer set semantics of logic program \mc{P} assigns to \mc{P} a collection of \emph{answer sets} --  interpretations  of \ground{\mc{P}}. An interpretation $I$ of \ground{\mc{P}} is an answer set for  \mc{P} if $I$ is minimal (w.r.t. set inclusion) among the  interpretations satisfying the rules of 
\[\begin{array}{ll}
\mc{P}^I = \{ A \la B_1, \ldots, B_m | & A \la B_1, \ldots, B_m, \mynot B_{m+1}, \ldots, \mynot B_{n} \in \mc{P} \mbox{ and} \\
 & I(\mynot B_{m+1}, \ldots, \mynot B_{n}) = \mathit{true} \} 
\end{array}
\]

A logic program can have a unique, many or none answer set(s). Therefore, we show that programs with a particular characteristic are guaranteed to have unique answer set. 

\vspace{5pt}
\noindent \textbf{Acyclic Programs.} We say that a program is \emph{acyclic} when there is no cycle in the program.The acyclicity in the program  is guaranteed by the existence of a certain fixed assignment of natural numbers to atoms that is called a \emph{level mapping}.

A \emph{level mapping} for a program \mc{P} is a function 
\[l: \mc{B}_\mc{P} \ra \mathbf{N}\]
where $\mathbf{N}$ is the set of natural numbers and $\mc{B}_\mc{P}$ is the Herbrand base for \mc{P}.  We extend the definition of level mapping to a mapping from ground literals to natural numbers by setting $l(\mynot A) =  l(A)$. 

Let \mc{P} be a logic program and $l$ be a level mapping for \mc{P}. \mc{P} is \emph{acyclic with respect to l} if for every clause $A \la B_1, \ldots, B_m, \mynot B_{m+1}, \ldots, \mynot B_n$ in \ground{\mc{P}} we find  
\[
l(A) > l(B_i) \ \ \textrm{for all $i$ with $1 \leq i \leq n$}
\] \mc{P} is \emph{acyclic} if it is acyclic with respect to some level mapping. 

Acyclic programs are guaranteed to have unique answer sets \cite{Baral2003}.


\subsection{XACML Semantics Based On ASP Semantics}
We can see from Sect.~\ref{s:mapping} that all of the XACML 3.0 transformation programs are acyclic. Thus, it is guaranteed that  $\mc{P}_{XACML}$ has unique answer set. 

\begin{proposition}
\label{prop:xacml-asp}
Let $\mc{P}_{XACML}$ be a program obtained from  XACML 3.0 element transformations and let $\mc{P}_{Q}$ be a program transformation of \Request \mc{Q}.  Let $I$ be the answer set of $\mc{P}_{XACML} \cup \mc{P}_{Q}$. Then the following equation holds
\[ \semantics{X}(\mc{Q}) = V \mbox{ iff } \val(X,V) \in I\]
\end{proposition}

\subsection{Analysis on Incompleteness Policies}
A set of policies is \textit{complete} if it always returns a decision given for any request. XACML  defines that there is  one \PS as the root of a set of policies. Therefore,  we formally express \textit{completeness property} as follows:
\[
\textrm{\textbf{complete}: } \forall Q: \semantics{PS_{root}}(Q) \neq \na
\]
where $Q$ for \Request and $PS_{root}$ is the root of \PS element in the set of policies.

We say that there is a \textit{gap} in the policy set if it is not complete. Hence, we formally express \textit{gap property} as follows:

\[
\textrm{\textbf{gap}: } \neg complete
\]
It is equal to 
\[
\textrm{\textbf{gap}: } \exists Q: \semantics{PS_{root}}(Q) = \na
\]

The idea of having gap property is to have  a logic program that can show answer sets whenever there is gap in the policies. We use the answer sets as the witnesses of the incompleteness policies. 

In order to check gap property we should generate all possible values restored in the database for each attribute. Each attribute only possible to have one value. Thus, we use \emph{cardinality constraint}  \cite{Simons2002,Syrjanen} and the encoding is as follows:
\prog{\mc{P}_{generate\_one}:\\
1 \{ subject(X) : subject\_db(X)\} 1 & \la \top.   \\
1 \{ action(X) : action\_db(X) \} 1 & \la \top. \\
1 \{ resource(X) : resource\_db(X) \} 1 &\la \top.  \\
1 \{ environment(X) : environment\_db(X) \} 1 & \la \top. \\
}
The intuitive meaning of the above cardinality constrains is that, for each subject in the database, exactly one instance of subject request is generated. The conversion holds for other attributes. 

We say there is a gap whenever we can find a request that makes value of the $PS_{root}$ is \na.  Here is the encoding:
\prog{\mc{P}_{gap}:\\
\gap & \la \val(PS_{root}, \na). \\
\bot & \la \mynot \gap. 
}
We force ASP solver to find a gap by putting a constraint $\bot \la \mynot \gap$. 

The answer sets  of program $\mc{P} = \mc{P}_{XACML} \cup \mc{P}_{generate\_one} \cup \mc{P}_{gap}$ are the witnesses that the set of policies encoded in $\mc{P}_{XACML} $ is incomplete. When there is no model satisfies the program then we are sure that the set of policies captures all of possible cases. 


\subsection{Analysis on Conflicting Policies}  
A conflict never occurs in XACML because the structure of policies where there is only one \PS as the root of all of policies and all of others policies are combined by combining algorithm. Each combining algorithm returns a single decision either permit or deny and never return both decisions in the same time. However, it is still interesting to know whether there are two \Rule  give conflict decisions.  We formally define a conflict is as follows:
\[
\textrm{\textbf{conflict}: } \exists Q: \semantics{R}(Q) = \p \wedge \semantics{R'}(Q) = \d
\]
In order to compute whether there is a conflict in the set of policies, we encode a logic program for conflict property as follows:
\prog{\mc{P}_{\mathit{conflict}}: \\
\conflict & \la \val(R, \p), \val(R', \d), R \neq R'. \\
\bot & \la \mynot \conflict. 
}
The same as gap condition, we force ASP solver to find a conflict by putting a constraint $\bot \la \mynot \conflict$. 

A conflict can be analysis whenever $\mc{P} = \mc{P}_{XACML} \cup \mc{P}_{generate\_one} \cup \mc{P}_{\mathit{conflict}}$ returns answer sets. The returning models are evidences where the conflict between \Rule occurs. We conclude that a set of policies is conflict-free if and only if program \mc{P} is unsatisfied, i.e., there is no returned model. 

\subsection{Analysis on Reachability Policies}
A policy is reachable if there is a request such that the decision is made based on this policy.  Usually in a big set of policies, there is a policy that is not reachable. This happens because policies are built based on several components and combined together.  We formally define a reachability property as follows:
\[
\textrm{\textbf{reachable(R)}: } \exists Q: \semantics{R}(Q) \neq \na. 
\]
where $Q$ is \Request element and $R$ is \Rule element. 

The encoding of reachability property in logic program is as follows:
\prog{ \mc{P}_{reachable}:\\
\reachable(R) & \la \val(R, E), E \neq \na. \\
}

Formally, a policy is not reachable if for every request either:
\begin{enumerate}
\item It always return \na.
\[\textrm{\textbf{unreachable(R)}: } \forall Q: \semantics{R} = \na\]
\item in case of permit-overrides combining algorithm, a policy is not reachable if its decision is deny but the final decision of the root policy is permit.
\[\textrm{\textbf{unreachable}: } \forall Q: \semantics{R} = \d \wedge \semantics{P}(Q) = \p\]
where in $P = [T, \seq{\ldots, R,\ldots}, \po]$
\item In case of deny-overrides combining algorithm, a policy is not reachable if its decision is permit but the final decision of the root policy is deny.
\[\textrm{\textbf{unreachable(R)}: } \forall Q: \semantics{R} = \p \wedge \semantics{P}(Q) = \d\]
where in $P = [T, \seq{\ldots, R,\ldots}, \denyo]$
\item In case of only-one-applicable combining algorithm, a policy is not reachable if it is applicable policy but the final decision of the root policy is not applicable. This indicates that there is another policy that is also applicable. 
\[\textrm{\textbf{unreachable(R)}: } \forall Q: \semantics{R} \neq \na \wedge \semantics{P}(Q) = \na\]
where in $P = [T, \seq{\ldots, R,\ldots}, \ooa]$

\item In case of first-applicable combining algorithm, a policy is not reachable if it is applicable but there is another policy in the same collection that is in the earlier of the sequence that is also applicable.
\[\textrm{\textbf{unreachable($R_j$)}: } \forall Q: \semantics{R_j} \neq \na \wedge \semantics{R_i}(Q) \neq \na \wedge i < j \]

where in $P = [T, \seq{\ldots, R_i, \ldots, R_j, \ldots}, \fa]$
\end{enumerate}

First of all we should generate all possible attributes. This time, the encoding is different with program $\mc{P}_{generate\_one}$ because we want to generate all possible attributes, not only one. Hence, we do not use cardinality constraint in this encoding. Here is the encoding:
\prog{\mc{P}_{generate\_all}:\\
subject(X) & \la subject\_db(X). \\
action(X) & \la action\_db(X). \\
resource(X) & \la resource\_db(X).\\
environment(X) & \la environment\_db(X).
}

Following we translate each unreachable condition into logic program
\prog{\mc{P}_{not\_reachable}:\\
\notreachable(R) & \la \mynot \reachable(R). \\
\notreachable(R) & \la \val(P, \p), \dec(P, R, \d). \\ 
\notreachable(R) & \la \val(P, \d), \dec(P, R, \p). \\ 
\notreachable(R) & \la \val(P, \na), \dec(P, R, E), E \neq \na.\\
}

In the case of first-applicable combining algorithm, there is a possibility a policy returns permit and the final decision is also permit, but, the permit of the final decision comes from the earlier applicable policy. Hence, we should take care of the ordering of policies.  We need to add extra rules in the program transformations such as: for every \Rule $R_i$ contained in $P_{id} = [T, \seq{R_1, \ldots, R_n}, \fa]$, $\mc{P}_{P_{id}}$ also contains:
\prog{
\dec(P_{id}, R_i, E, I) & \la \val(R_i, E). \ (1 \leq i \leq n)
}

Here we add to our $\mc{P}_{not\_reachable}$
\prog{
\notreachable(R_j) & \la \dec(P_{id}, R_i, E, I), \dec(P_{id}, R_j, E', J), E \neq \na, E' \neq \na, I < J.\\
}

To check unreachable property we add to our program $\mc{P}_{not\_reachable}:$
\prog{
\notreachable & \la \notreachable(R). \\
\bot & \la \mynot \notreachable. 
}
We force ASP solver to find unreachable policies by putting a constraint $\bot \la \mynot \notreachable$. When $\notreachable(R)$ is in the answer set of $\mc{P} = \mc{P}_{XACML} \cup \mc{P}_{generate\_all} \cup \mc{P}_{reachable} \cup \mc{P}_{not\_reachable}$ then it is safe to remove policy $R$ from the set because it is unreachable.  
\section{Conclusion and Future Work}
We have shown a mechanism to map  XACML 3.0 components into logic programs. Using the advantages of ASP technique to solve combinatorial problems efficiently we have presented ASP programs to capture  analysing in access control policies incompleteness  property, conflicting property and unreachability property. Our intention is to have an automatic tool that both showing formalization and also help policy administrators to analyse their policies sets.

For future work, we would like to analyse conflict in attribute based like in Singh's work \cite{Singh2010}. We also would like to extend our work to handle Role-Based Access Control (RBAC) \cite{xacml3.0rbac} and see the conflict might occurs between different roles. 

In order to reduce the policies, we could inspect redundancy between policies. We should find a subset of policies that might capture the whole possible decisions might happen in all policies. Thus, we could have smaller set than the original policies set. 

\bibliographystyle{plain}
\bibliography{bibliography}

\begin{thebibliography}{10}

\bibitem{Baral2003}
Chitta Baral.
\newblock {\em Knowledge Representation, Reasoning and Declarative Problem
  Solving}.
\newblock Cambridge University Press, February 2003.

\bibitem{Belnap1977}
N.D. Belnap.
\newblock A useful four-valued logic.
\newblock In G.~Epstein and J.M. Dunn, editors, {\em Modern Uses of
  Multiple-Valued Logic}, pages 8--37. D. Reidel, Dordrecht, 1977.

\bibitem{Bruns2008}
Glenn Bruns and Michael Huth.
\newblock Access-control via belnap logic: Effective and efficient composition
  and analysis.
\newblock In {\em 21st IEEE Computer Security Foundations Symposium}, June
  2008.

\bibitem{Gelfond2007}
Michael Gelfond.
\newblock Handbook of knowledge representation.
\newblock In B.~Porter F.~van Harmelen, V.~Lifschitz, editor, {\em Foundations
  of Artificial Intelligence}, volume~3, chapter Answer Sets, pages 285--316.
  Elsevier, 2007.

\bibitem{XACML3.0}
Erik Rissanen.
\newblock e{X}tensible {A}ccess {C}ontrol {M}arkup {L}anguage ({XACML}) version
  3.0 (committe specification 01).
\newblock Technical report, OASIS,
  http://docs.oasis-open.org/xacml/3.0/xacml-3.0-core-spec-cd-03-en.pdf, August
  2010.

\bibitem{xacml3.0rbac}
Erik Rissanen.
\newblock Xacml v3.0 core and hierarchical role based access control (rbac)
  profile version 1.0 (committe specification 01).
\newblock Technical report, OASIS,
  http://docs.oasis-open.org/xacml/3.0/xacml-3.0-rbac-v1-spec-cs-01-en.pdf,
  August 2010.

\bibitem{Samarati2001}
Pierangela Samarati and Sabrina de~Capitani~di Vimercati.
\newblock Access control: Policies, models, and mechanisms.
\newblock In {\em Foundations of Security Analysis and Design, Tutorial
  Lectures}, volume 2171 of {\em Lecture Notes in Computer Science}, pages
  137--196. Springer Verlag, 2001.

\bibitem{Simons2002}
Patrik Simons, Ilkka Niemel\'{a}, and Timo Soininen.
\newblock Extending and implementing the stable model semantics.
\newblock {\em Artificial Intelligence}, 138(1-2):181--234, 2002.

\bibitem{Singh2010}
Kamalbir Singh and Sarbjeet Singh.
\newblock Design and evaluation of {XACML} conflict policies detection
  mechanism.
\newblock {\em International Journal of Computer Science and Information
  Technology}, 2:65--74, 2010.

\bibitem{Syrjanen}
Tommi Syrj{\"a}nen.
\newblock {\em Lparse 1.0 User's Manual}.

\end{thebibliography}
\end{document}